\documentclass[twocolumn,showpacs,preprintnumbers]{revtex4-1}

\usepackage{graphicx}
\usepackage{color}
\usepackage{longtable}

\begin{document}

\title{Study of the hidden-order of URu$_{2}$Si$_{2}$  by point contact
tunnel junctions.}

\author{R. Escudero }
\email[Author to whom correspondence should be addressed. R. Escudero: ]
{escu@unam.mx} \affiliation {Instituto de Investigaciones en Materiales,
Universidad Nacional Aut\'{o}noma de M\'{e}xico. A. Postal 70-360.
M\'{e}xico, D.F. 04510 MEXICO.}

\author {R. E. L\'{o}pez-Romero}
\affiliation {Instituto de Investigaciones en Materiales, Universidad
Nacional Aut\'{o}noma de M\'{e}xico. A. Postal 70-360. M\'{e}xico, D.F.
04510 MEXICO.}
\author{F. Morales.}
\affiliation {Instituto de Investigaciones en Materiales, Universidad
Nacional Aut\'{o}noma de M\'{e}xico. A. Postal 70-360. M\'{e}xico, D.F.
04510 MEXICO.}

 \date{\today}

\begin{abstract}
{ URu$_{2}$Si$_{2}$ presents superconductivity at temperatures below 1.5
K, and a hidden order (HO) at about  17.5 K,  both electronic phenomena
are influenced  by  Fano and Kondo resonances. At 17.5 K the HO was
related in the past to a Peierls distortion that produces an  energy  gap
deformed  by the resonances. This  order  has been studied by more than
20 years, still without a clear understanding. In this work we studied
the  electronic characteristics of  URu$_{2}$Si$_{2}$} in  a single
crystal, with  tunneling and metallic point contact spectroscopies. In
the superconducting state,  we determined the energy gap, which  shows
the influence of the Fano and Kondo resonances. At temperatures where HO
is observed,  the tunnel junctions spectra show   the influence of the
two resonances. Tunnel junctions  characteristics  show that the Fermi
surface nesting depends on the crystallographic  direction.
\end{abstract}

\maketitle

\section{Introduction}

URu$_{2}$Si$_{2}$, is one of the most studied materials, after more  than
20 years the electronic characteristics  are still not well understood.
One notable example is the so called hidden order which remains
unexplained. Transport measurements in this compound have  shown many
details related with the general behavior; from room temperature to 70 K
URu$_{2}$Si$_{2}$ presents a characteristic typical of a  Kondo  system
associated to the interaction of heavy $f$ electrons and $spd$ conducting
electrons. With a  Kondo lattice formed at $T_K \sim$ 70 K
\cite{haule09,chatterjee13},  below this temperature the resistivity
dramatically  drops, and at $T = 17.5$ K an anomaly is observed. In the
past this was related to a spin density wave (SDW). However this  feature
is called the Hidden Order (HO).  It presents many diverse  electronic
characteristics that modifies other occurring proceses
\cite{L.Buyers,P.Chandra, Oh, mydosh}. At lower temperature  below $\sim$
1.5 K the  compound, turns into a superconducting state  which many
researchers believe must be anomalous in the sense that does not follow
the BCS model \cite{maple, ramirez, ohkuni, kwok, moyoyama, hassinger}.

 As mentioned before  the HO was associated in the past to a
Peierls distortion, but careful experiments carried out to detect the
Peierls (SDW) distortion  did not find any  magnetic order, only a quite
small staggering magnetism, probably related to  magnetic impurities, not
related to  SDW behavior. This absences imply that the gaping on the
Fermi surface may be related to another unknown electronic phenomena
\cite {L.Buyers, Broholm}. Actually, doubts still persist about the
physics behind the anomalies at about  17.5 K. The HO changes  transport,
and thermal properties, and opens an energy gap, and  altering the
superconducting behavior  \cite{Oh, M.Palstra, mydosh, ramirez, deVisser,
U.Walter, Mason}.
 The occurrences of the  HO, and the  relationship with the  Fano and
 Kondo resonances is  not completely understood, nobody  knows the
 influence of the two resonances and how those are affecting the general
 behavior, for instances the manner that the Kondo lattice at high
 temperature and the resulting distortion at  17 K modifies the
 electronic characteristics.
Experiments with very sensitive techniques as tunneling and  point
contact spectroscopy (PCS) provide the observation of the Fano resonance
that distorted the feature of the energy gap,  at $T\sim 17.5$ K as
observed by Schmidt, et al, with Scanning Tunneling Spectroscopy \cite
{schmidt}.

Many researchers who investigated this transition demonstrated that this
is  a  second order thermodynamic transition \cite{maple,ohkuni,
escudero, morales,schlabitz}.  The  effects of the resonances   at the
transition show the consequence of a partial electronic screening that
modifies the transport and thermal  properties  because   the   strong
hybridization between $spd$ conducting  electrons and heavy localized $f$
electrons \cite{schmidt,aynajian}.
 The result  of all those processes is that the Kondo and Fano lattices
 are spatially modulates.
 The signature of  these lattices  are  displayed as modifications at the
 Fermi surface. Thus,  portions of the Fermi surface at high and low
 temperatures are distorted and observed   with  many experimental tests
 and theoretical studies \cite{haule09,maple,villaume,chandra,rodrigo}.

This work  shows our new experimental observations obtained with
techniques that  give additional information  about the  features and
characteristics of this compound using tunneling and PCS \cite{IvarG,
IvarGi, K.Yanson, A.Duif, Jansen}.

The studies at high temperatures from 40 to 1.7 K were performed using
tunnel junctions to observe the HO  that develops at $\sim$17.5 K. We
studied the anomalous features of the  gap, using a well characterized
URu$_{2}$Si$_{2}$ single crystal \cite{escudero,morales,hass},  we
measured the spectroscopic  characteristics of the  crystal in
two crystallographic directions.

At lower temperature the superconducting state was studied with metallic
point contacts (PCS) in the range of temperatures where superconductivity
develops, the studies were performed from   2.5  to 0.3 K.  PCS junctions
were in the diffusive limit but close to the ballistic  regime.  In all
experiments, here reported,   from 40 K to 0.3 K  the influence of the
Fano and Kondo resonances was observed. In the superconducting state the
influence of both resonances was clearly evidenced; the   feature of the
energy gap was completely distorted, and  the effect of these resonances
is clearly noted when examined  the evolution with temperature of the
energy gap.

 In the range of temperatures where HO is set, the influence was studied
 in two directions of the crystal, the Fano features are different in
 both directions.

  This work  reports tunneling and point contact experiments performed in
  URu$_{2}$Si$_{2}$ single crystals. At temperatures from 1.7 K to 40 K,
  tunneling spectroscopy  shows  the effect of the Fano and Kondo
  resonances on the energy gap, and the dependence on the
  crystallographic  direction.  The energy gap in the $a$ direction is
  bigger than the gap in the $c$ direction. In the superconducting state,
  with PCS  we found that the energy gap as a function of temperature
  follows the BCS prediction after the Fano anomaly is subtracted from
  the differential conductance.

\section{Experimental details}
   The URu$_{2}$Si$_{2}$ single crystals used in this study were  growth
   by Czochralski method and annealed for one week  at 850$^\circ$ C.
   They have similar characteristics  to that reported elsewhere
   \cite{hass, morales, escudero}. The crystals have a platelet-like
   shape with the crystallographic ${\bf c}$ axis perpendicular to the
   plane of the platelet, typical dimension of the used crystals were
   about 2x3x0.5 mm$^{3}$.

The samples were characterized by resistance as a function of temperature
measurements. The applied  current was  100 $\mu$A. The resistance ratio
between room temperature and low temperature resistance R(300 K)/R(2
K)=38. The anomaly associated to the HO is observed clearly at about 17.5 K. At lower temperature, the
superconducting transition temperature is at $T_C$=1.37 K (R=0), this
transition is also similar to the observed in other experiments by other
researchers. The  transition temperature width was  0.15 K.
   Two type of junctions were formed, tunnel
   junctions  were build  by using a tip of Au(W) wire  and the
   URu$_{2}$Si$_{2}$  crystal. The
   insulating barrier that forms the tunnel junction was the native
   oxide
   on the surface of the compound, and/or the oxide on the tip of the
   wire.
    The reason to assume that the junctions in our experiments behaves as tunnel or point contacts is the oxide formed in the wire or the sample, two considerations we have for this behavior; the shape of the differential conductance - bias voltage shows a  typical curve of a tunnel junction \cite{akerman}. In tunnel junctions  the zero bias minimum in the curves is displaced out of the origin. Aspect quite different to a metallic contact junction, where the minimum in the differential conductance always occurs at zero bias voltage  \cite{akerman}. However another important
   consideration is the value of the differential resistance at zero bias, in our tunnel junctions, this is of the order about 20 - 50 $\Omega$.
    Several
   try outs  were performed until reproducible data was obtained.
   Junctions with differential resistances at zero bias voltage between
   20$-$50 $\Omega$ gives the most reproducible data. The two
   crystallographic directions of single crystal were measured with  the
   prepared  tunnel junctions;  one used to measure over the edge of the
crystal axis, the  ${\bf a}$ direction,  and the another on the plane of
the platelet ${\bf c}$ axis.
   For the measurements and characterization of the superconducting
   state,  point contact junctions were used. These
   were formed with the single crystal and a fine tungsten-gold tip with
   diameter on the tip, smaller than  5 $\mu$m, those  junctions were
   prepared at room temperature and glued to a glass substrate with
   Oxford varnish.  The area of the junctions were estimated to be $\sim$ 1 $\mu$m$^2$.

The differential resistance d$V$/d$I$ as  a function of the bias voltage
$V$ of the point contacts  was measured with the standard modulation ac
lock-in amplifier technique, and bridge.
 The temperature range for characterization of the superconducting state
 was from 0.325 to about 3 K. For characterization of the hidden order
 the temperatures range was from
1.7 K to  40 K using a MPMS system from Quantum Design as cryostat,
whereas  below 2 K  a  $^{3}$He Oxford refrigerator was used.

The characterization of the work regime of the point contacts  was
estimated with the Wexler's interpolation formula \cite{wexler}; we
substituted the mean free electronic path $l_{i}\simeq 100$ \AA ~
\cite{hass}, the resistivity $\rho \sim $ 40 $\mu \Omega $-cm measured at
2 K \cite{rauch} and the resistance of the point contacts measured at
zero bias voltage. The obtained radii values,  between 320 \AA and 3700
\AA, indicate that the contacts are in the diffuse regime, but not far to
the ballistic \cite{duif}.

An important aspect of the junctions used was the thermal stability. In
this work we report studies performed in more than 20 different contact
junctions, all  shown similar $dV/dI(V)$ curves.

 \begin{figure}
\centerline{\includegraphics[scale=0.34]{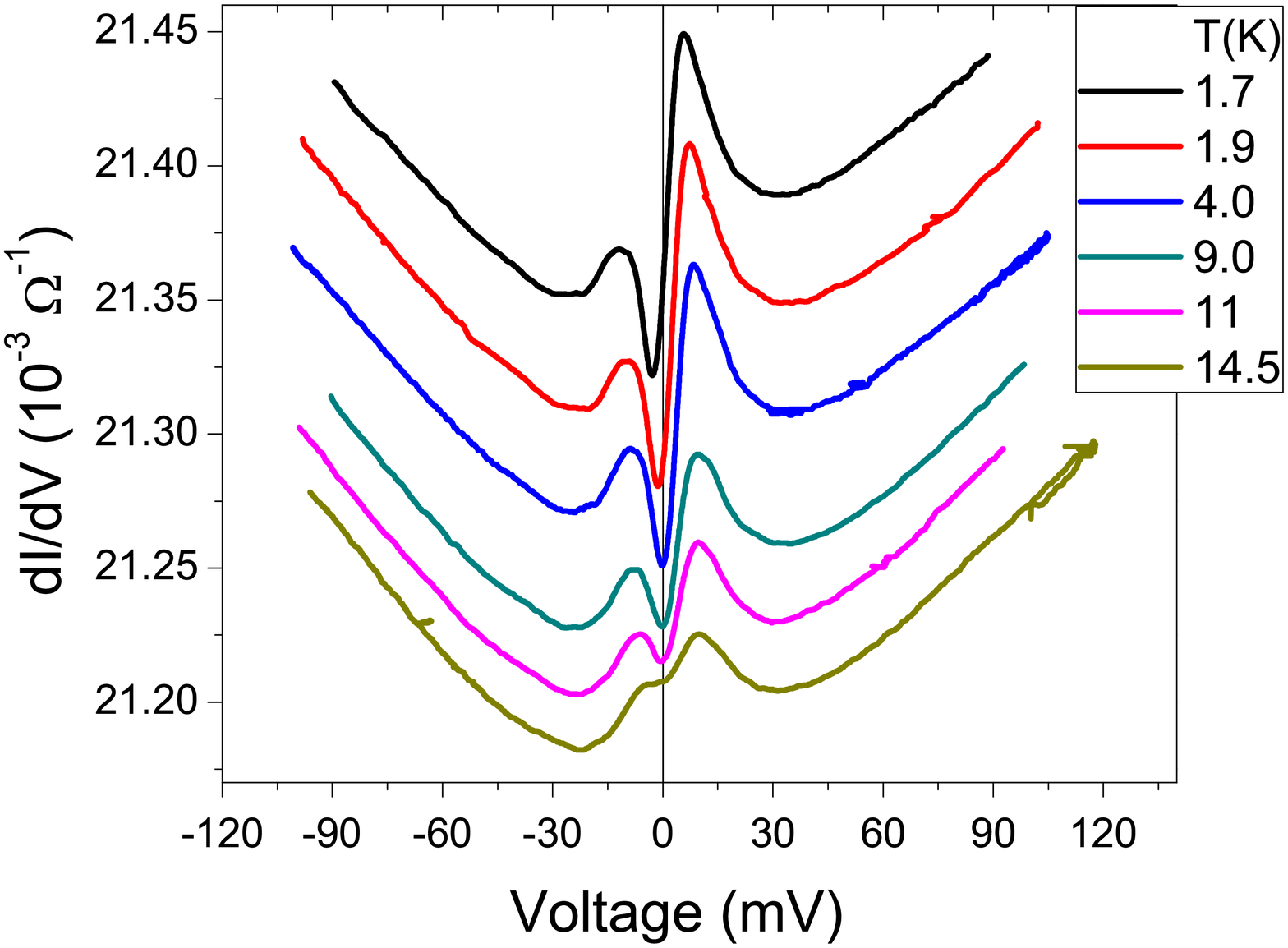}} \caption{(Color
online)  Differential conductance as a function of voltage of
URu$_2$S$_2-$Au(W) tunnel junction, at temperatures from 1.7 K to 14 K.
The differential resistance  at zero bias is about $\sim$46 Ohm. Note the
sharp characteristics and the smoothing as the temperature rises. The
parabolic background at high bias is typical of a  normal tunnel
junction. Only the characteristic between $\pm$ 30 mV presents the Fano
resonance, the curves were displaced vertically for better clarity.}
\label{Fig.1}
\end{figure}

\begin{figure}
\centerline{\includegraphics[scale=0.34]{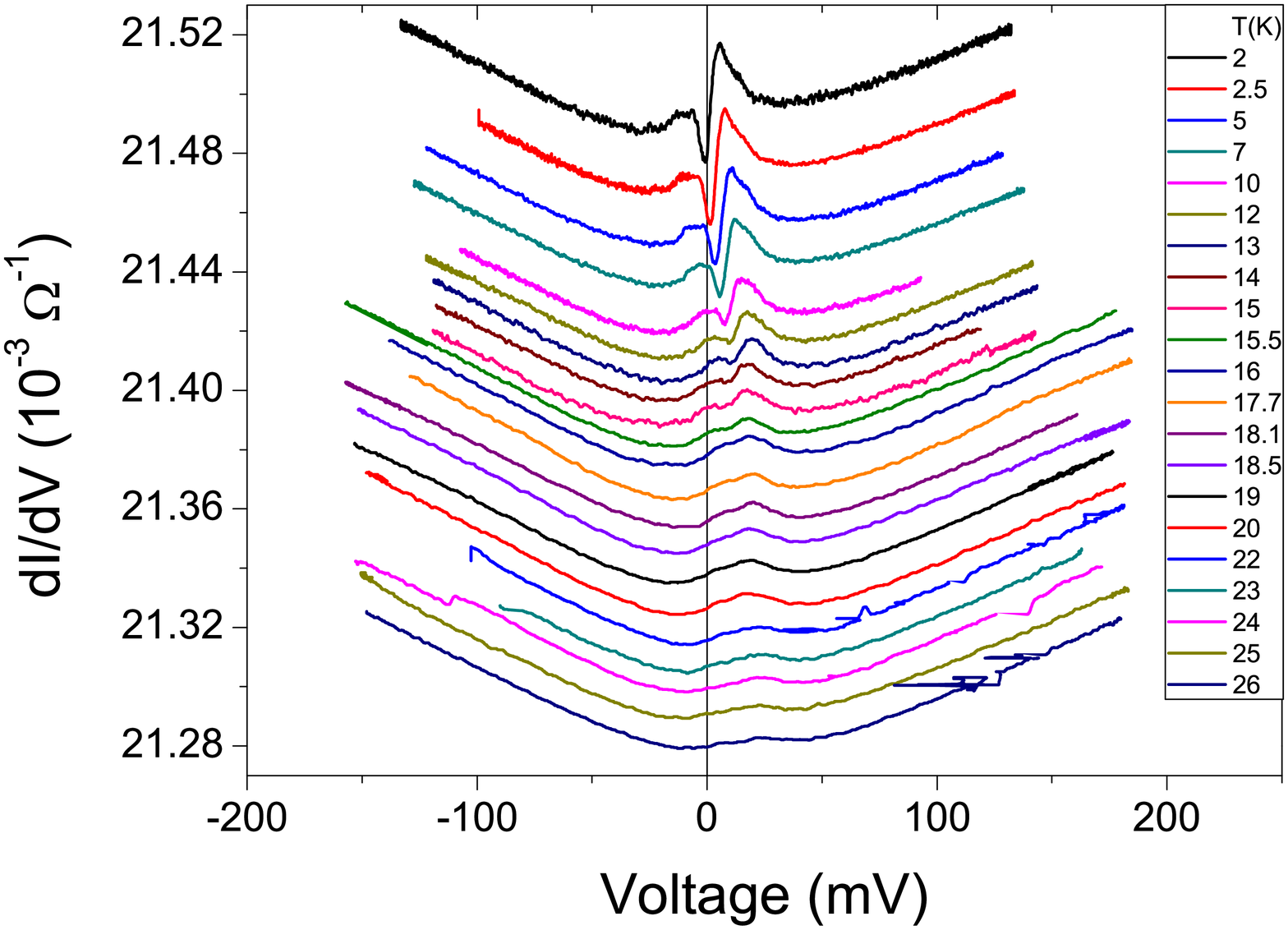}} \caption{(Color
online) Differential conductance of a tunnel junction in the $c$
direction taken from  2 to 26 K. Note the reduction of the predominant
sharp characteristics close to the Fermi level when the temperature is
increased.  The minimum is displaced from  the origin  to positive bias
voltage. Raising the temperature the  structure is smoothed.   At high
bias voltages  the background of the tunnel junction looks normal. All
data was vertically displaced for a clear view.} \label{fig.2}
\end{figure}

\begin{figure}
\centerline{\includegraphics[scale=0.34]{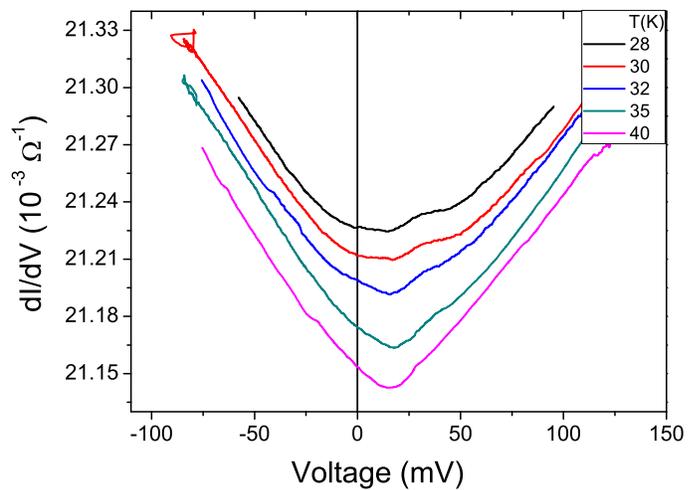}} \caption{(Color
online) Data for another tunnel junction at high temperature from 28 to
40 K. Here also it is possible to see that small distortion still
persist. The minimum  observed in the figure 2 now clearly is at about 18
mV. However the distortion of the Fano resonance is difficult to see, at
40 K the characteristic is of a typical normal tunnel junction, if some
anomalies persist are difficult to see.} \label{fig.3}
\end{figure}

\section{Results and discussion}

Temperature evolution of the differential conductance versus bias voltage
from 1.7 to  14.5 K are presented in Fig.1. These curves reveal the
spectroscopic features of the Fano and Kondo resonances, measured in the
$c$ crystalline direction. The  spectroscopic features of the Fermi
surface and the density of states are strongly modified by the Fano and
Kondo lattices. Note that in a Peierls distortion the effects are quite
different.

The evolution of the spectroscopic features,  when  temperature is
increased, decreases the distortion in the differential conductance. Only
close to T = 17.5 K where the hidden order develops the distortion is
very small.
 Fig. 2 shows measures  performed from 1.7 K to 26 K, in ${\bf c}$
 direction of the
crystal, the distortion decreases as the temperature increases, but it
persists at high temperature,  the curves measured of the differential
conductance characteristic at 1.7 K show a typical Fano resonance shape,
with very sharp structure. Those characteristics were also observed with
similar detail by Elgazzar, et al. and Aynajian, et al.
\cite{aynajian,elga}. One set of measurements in the same direction,
${\bf c}$,  up to 40 K,  shown in  Fig. 3, displays  the typical
parabolic background of  a  normal tunnel junction,  with  the
conductance minimum shifted from the zero bias voltage, out of the
origin \cite{wolf}.  However, close  examination of the curves still
indicates that some features persist  at bias voltages at  25$-$45 mV.
Those features may be related to  the Kondo lattice.

Fig. 3  shows  the smoothing of the central peak and the displacement of
the minimum  out of the origin. However, also  the normal parabolic shape
of a tunnel junction is noted, but with very small structure  \cite{wolf}
perhaps because the remanence of the  Kondo lattice.   Summarizing, Figure 2 and 3 shows the effect of temperature on the features around zero bias that we are attributed to the formation of a possible arising of
a hybridization gap by the Kondo lattice at temperatures for above 70 K \cite{park12}.
In Figure 4 we show data measured  in the $a$ crystallographical
direction, these data measured from  20 to 2 K. The spectra are
different to the measurements in $c$ direction,  in  the $a$ direction
the differential conductance shows two peaks at $\pm 10$ mV, which  tends
to decrease  as the temperature was increased.  At 20 K there is no
distortion.      Note that the Fano anomaly looks completely different to
the  observed in the $c$ direction. The  data was measured and was
reproducible when measured in  different junctions.  This Fano distortion
is different to as the Fano theoretical  model \cite{fano}.  Figure 4
shows two peaks in the differential conductance,  these are quite
symmetrical  with similar magnitude.  Those peaks observed at  2 K, are
separated from the  origin   at  $\pm$ 10 mV,  we may consider that the
difference of voltage between the peaks  may be the size of an  energy
"gap" of the HO, big, as  the gaping in  $c$ direction \cite{maple}.
Similarly  as in other figures and directions,  figure 4 shows a
decreasing structure, tending to be smoothed  as temperature is
increased. This structure  disappears at 20 K.  We noted that in ${\bf
a}$ direction no structure exist at 20 K, but in ${\bf c}$ direction
still remains. This is in agreement with the observations by Haule and Kotliar in U or Si site  \cite{haule09}.  Also  we have to  mention that in total accord with  Park  {\sl et al.} \cite{park12} the new features of the Fano resonance, as observed in Fig. 4 with measurements taken in the $a$ crystallographic direction of our sample, clearly  may indicate  another type of a  Fano resonance  and/or  feature of a hybridization gap.

Our experiments, show that  nesting depends on the crystallographic
direction, and this may be related  to
 size of the HO  gap. Measurements  in $a$ direction show that
 2$\Delta$= 20 mV, which is
small in $c$, as already determined by Aynajian et al. as
  $\sim$8 mV \cite{aynajian}.
With the analysis of the characteristics at temperatures from 40 to 1.7
K, measured in the $a$ direction is possible to do a determination of the
HO  gap; at 2 K the two maxima are $\pm$10 mV, and the ratio
2$\Delta$/K$_B$T$_C$ is $\sim$ 13.

Point contact experiments reported by Naidyuk {\sl et al.}
\cite{naidyuk,naidyuk96}, from temperatures above the
superconducting transition up to 20 K, show the differential conductance
as a function of bias voltage measured in the $c$ direction and
perpendicular to it. In those experiments the features observed in the
$c$ direction are not show, meanwhile in the $a$ direction an energy gap
is observed with a value about 10 mV, but  no features were observed
above 17.5 K. Our experiments reported here show features up to 40 K. As mentioned above this  behavior over 17.5 K might be
interpreted as the arising of a hybridization gap by normalized bands
\cite{park12}. Our measurements, in the $c$ direction, display more
structures which may be associated to an energy gap, persisting above 17.5 K up to  40 K in agreement to Park {\sl et al.} \cite{park12}.

\begin{figure}
\centerline{\includegraphics[scale=0.34]{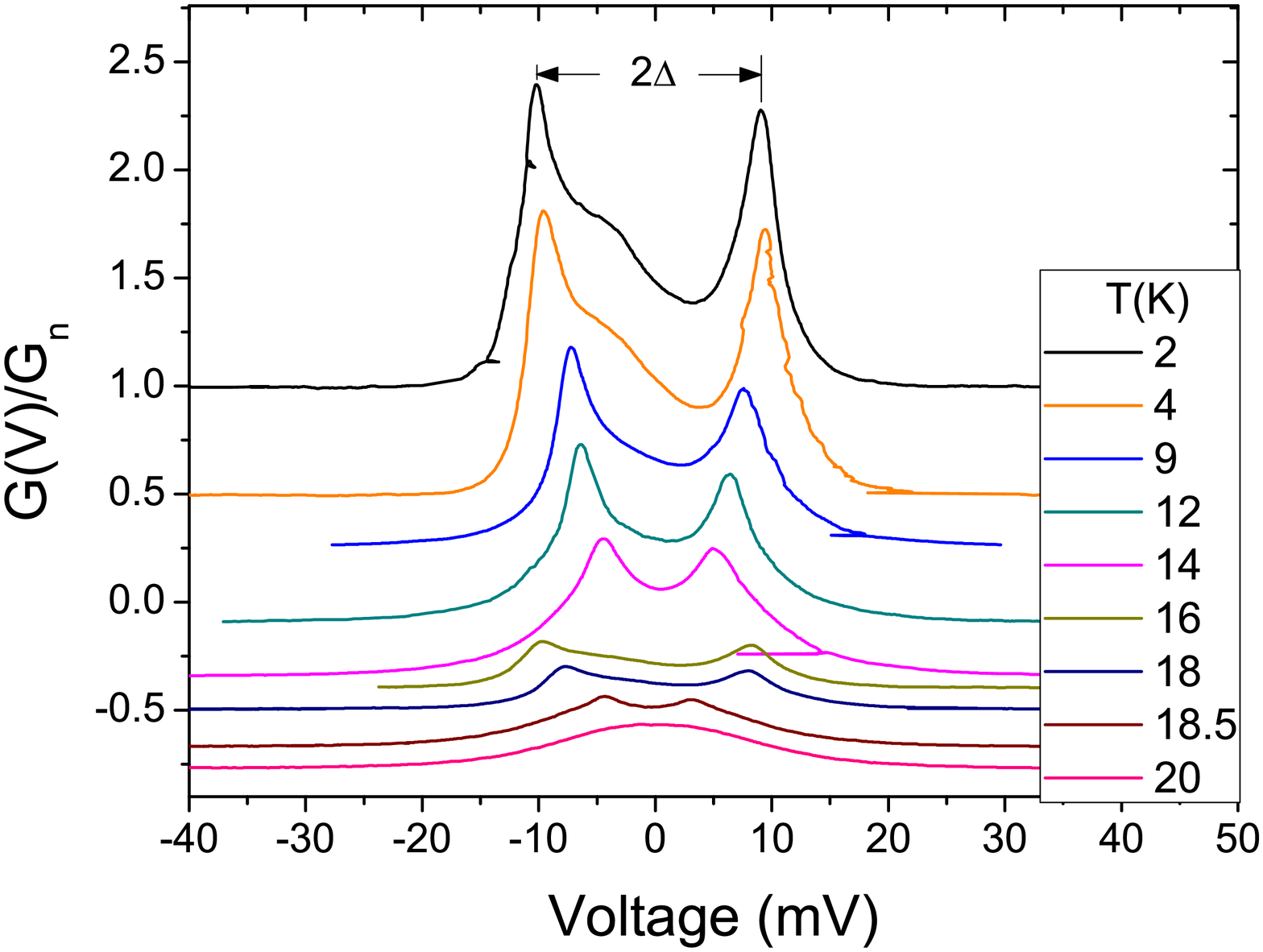}} \caption{(Color
online) Normalized differential conductance of a tunnel junction measured
in the $a$ crystallographic direction.  These curves show the presence of
a bigger distortion, different to the observed  in $c$ crystallographic
direction. Note that the  structure of the Fano resonance is distorted
with two symmetrical peaks at $\pm$ 10 mV.  The features decreases as the
temperature rises, at 14 K  the feature look similar to a contact
junction with a $Z$ parameter about 0.2 \cite{btk}.} \label{fig.4}
\end{figure}

Studies with PCS  were performed at low temperature in the
superconducting state to  see the evolution with temperature of  the
energy gap. The features  were characterized from 3 K to 0.325 K using a
$^3$He as cryostat. In  Figure 5A  we show the differential conductance
between 0.325 K and 1.7 K, all curves were displaced in the vertical
direction by a small amount  to have a clear view and details. At
temperatures above the superconducting state the Fano  resonance may be
observed as a distortion at  the central part of the curves at zero bias
voltage, this behavior was observed and analyzed   when plotting the
evolution of the gap with temperature. At the lower accessible
temperature the superconducting gap feature is well formed, in panel B of
figure 5 we display  and plot  temperature evolution of the gap.

\begin{figure}
\centerline{\includegraphics[scale=0.32]{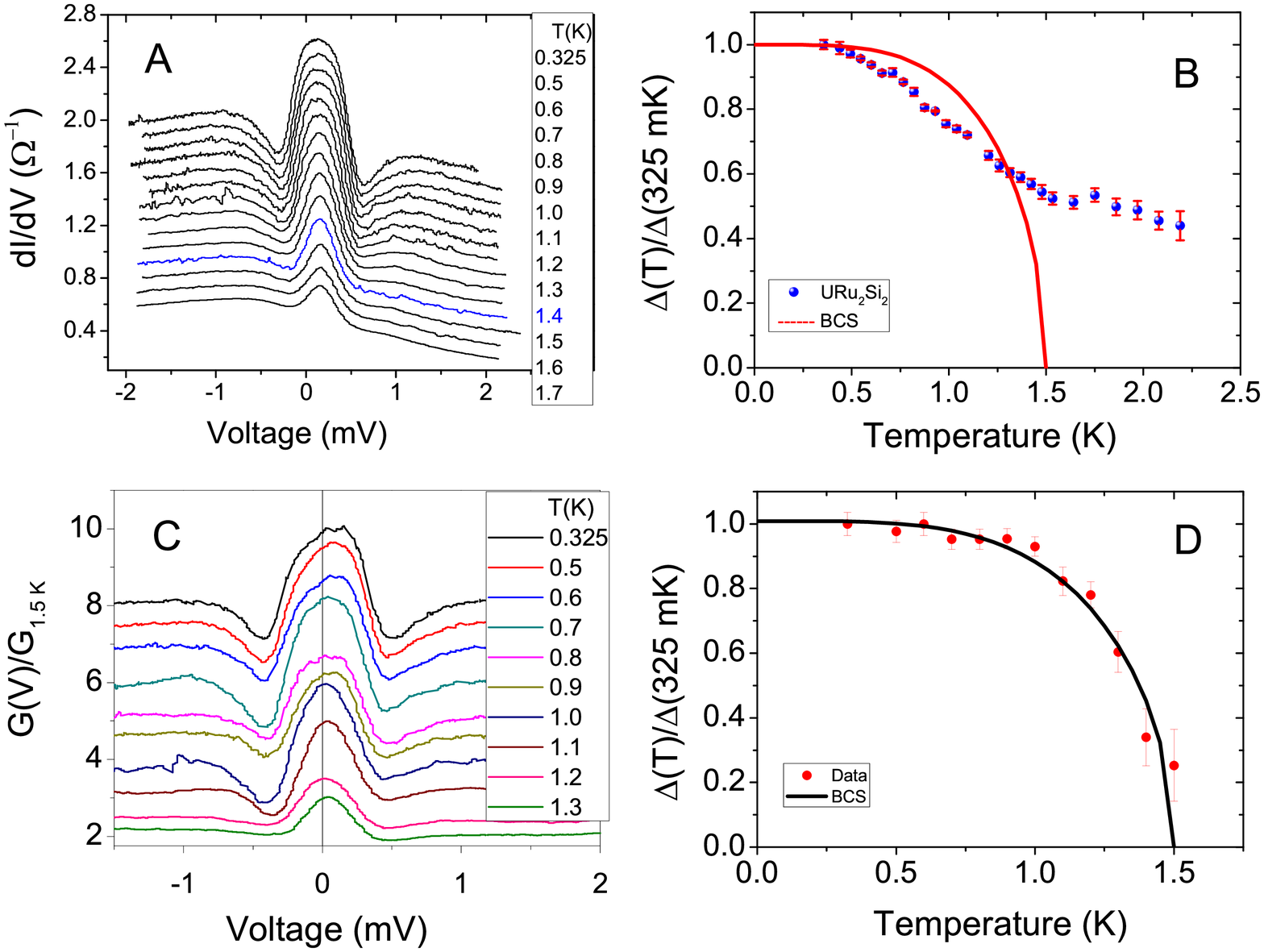}} \caption{(Color
online) Normalized differential conductance of a point contact in the
superconducting state;  URu$_2$Si$_2$ - Au(W). Panel A shows the rough
data measured from {\color{red} 1.7}  to 0.3 K. The gap value versus
temperature is displayed in B. The evolution does not follow the BCS
theory. In order to see the influence of the Fano resonance we subtracted
from these data the curve measured at 1.7 K, the resultant curves are
shown in panel C, there we plotted the differential conductance
normalized at 1.5 K. In D is plotted the superconducting energy gap as a
function of normalized temperature, this data follow quite well the BCS
theory. This strong and different evolution with temperature  of the gap
is due to the Fano resonance.} \label{fig.5}
\end{figure}

As show in Fig. \ref{fig.5}B the energy gap evolution does not follows the BCS model (see continuous line). This anomalous behavior \cite{morales} is
the effect of the Fano and Kondo resonances. In Fig. 5C it is presented
the  normalized differential conductance of the same PCS with the
structure related  to the Fano resonance subtracted  in  all data.   The
subtracted data was the differential conductance measured at 1.7 K.  The data substraction was performed in a similar manner as performed by Aynajian {\sl et al.} \cite{aynajian}.
Fig. 5 panel D presents the results of this procedure.
Data of the superconducting gap as a function of temperature now  clearly
follow BCS. The characteristics shown in Fig. 5B and 5D, show the
influence of the resonances in the evolution with temperature of the gap.
Our measurements also indicate that the ratio
2$\Delta$/K$_B$T$_C$ is  $\sim$7.86 -7.57. with  the transition
temperature between 1.35 - 1.4 K, this value is characteristic of  a
strong coupling superconductor.   However, it is very important to
mention that recent studies  by Kawasaki {\sl et al.}, and Schemm {\sl et al.} \cite{kawa,schemm} have shown time reversal symmetry breaking, using
different experimental techniques. Kawasaki, {\sl et al,} studied the HO and
superconductivity using muon spin relaxation spectroscopy, whereas
Schemm, {\sl et al.}  using  Kerr effect. 
Both experiments found evidence of
broken time reversal symmetry, which has a strong implication for both phenomena occurring in URu$_2$Si$_2$.  Some features below the superconducting state, performed with   polar Kerr experiments were observed and as suggested  by Schemm {\sl et al.} this effects may imply other physical processes in the superconducting state,  and well above  in the HO  phase at 25 K \cite{schemm}.  Therefore, as mentioned by Kawasaki {\sl et al.} and Schemm {\sl et al.}, those observations may imply a novel pairing mechanism for the superconducting behavior.

\section{Conclusions}

In summary, we have studied the electronic  characteristics  of
URu$_2$Si$_2$ with   tunnel and point contacts  spectroscopies. The spectra
show a  gap feature  with distortions because  Fano and Kondo resonances,
at high and low temperatures. At low temperature the superconducting gap
has an anomalous  evolution with temperature. From  0.3  to 2.4 K,  we
found the influence  of the Fano and Kondo resonances  and were
subtracted, then energy gap follows BCS. From 1.7 to 40 K we observed the
presence of the resonances that distorted  the features on the Fermi
surface, those look different in different crystallographic directions.
Lastly,  in the $c$ direction the  structure in the differential
conductance  persists  up to 40 K, whereas in the $a$ direction it
persists up to  20 K. The HO  effects are different in both directions of
the Fermi surface.

\begin{acknowledgments}
This work was partially supported by the Direcci\'{o}n General de Asuntos
del Personal Acad\'{e}mico, UNAM, project IN106414;   by Consejo Nacional
de Ciencia y Tecnolog\'{i}a, (CONACyT),  project 129293, (Ciencia
B\'{a}sica),  BISNANO of the European Community and M\'{e}xico, and
project PICCO 11-7 by the Instituto de Ciencias del Distrito Federal, Cd.
de M\'{e}xico.
\end{acknowledgments}

\end{document}